\documentclass[12pt]{article}

\usepackage{graphicx}
\usepackage{color}
\usepackage{epstopdf}
\usepackage{subfigure}
\usepackage{amssymb}
\usepackage{amsmath}
\usepackage[latin1]{inputenc}

\providecommand{\abs}[1]{\lvert#1\rvert}

\newcommand{\ceil}[1]{ \lceil #1 \rceil }

\title{ Plane-Sweep Incremental Algorithm: Computing Delaunay Tessellations of Large Datasets }
\author{
Márton Trencséni\footnote{trencseni@complex.elte.hu}, István Csabai \\
\emph{Dept. of Physics of Complex Systems,  } \\
\emph{Eötvös University, Pázmány Péter sétány 1/A} \\
\emph{1117 Budapest, Hungary} \\
}

\begin{document}

\maketitle

\begin{abstract}
We present the plane-sweep incremental algorithm, a hybrid approach for computing Delaunay tessellations of large point sets whose size exceeds the computer's main memory. This approach unites the simplicity of the incremental algorithms with the comparatively low memory requirements of plane-sweep approaches. The procedure is to first sort the point set along the first principal component and then to sequentially insert the points into the tessellation, essentially simulating a sweeping plane. The part of the tessellation that has been passed by the sweeping plane can be evicted from memory and written to disk, limiting the memory requirement of the program to the "thickness" of the data set along its first principal component. We implemented the algorithm and used it to compute the Delaunay tessellation and Voronoi partition of the Sloan Digital Sky Survey magnitude space consisting of 287 million points.
\end{abstract}

\section{ Introduction }

Delaunay tessellation is a standard problem of computational geometry \cite{Fortune}. Let $S$ be a set of points in $d$-dimensional Euclidean space $E^d$, called sites. The Delaunay tessellation $D(S)$ of $S$ is a tessellation such that no simplex' circumsphere contains elements of $S$; this is called the Delaunay condition. The edges of the tessellation are called Delaunay edges, the edges generate the Delaunay neighbors. The dual of Delaunay tessellation is Voronoi partitioning. The Voronoi partition $V(S)$ of $S$ is a decomposition of $E^d$ into convex cells, such that each cell contains one element $s$ of $S$ and for all points $x$ in this cell, $d(x, s) \leq d(x, s')$ for all other $s'$ in $S$. $D(S)$ and $V(S)$ are each other's dual in the sense that if $v_1$ and $v_2$ are neighboring cells of $V(S)$ containing $s_1$ and $s_2$ of $S$, then there is Delaunay edge between $s_1$ and $s_2$, and vice versa.

In terms of implementation, the Delaunay tessellation is first calculated, from which the Voronoi partition can easily be computed. Let $N(s)$ denote the Delaunay neighbors of $s$, then the cell containing $s$ has $\abs{N(s)}$ faces because each neighbor generates a face defined by the plane whose points are equidistant from $s$ and the neighbor.

Several algorithms for computing Delaunay tessellations have been proposed and many implementations exist. However, all are in-memory implementations, meaning that they build the entire tessellation using virtual memory retrieved from the operating system using malloc() or similar calls. Thus, the amount of memory a process can allocate limits the size $\abs{S}$ of the tessellation it can compute, e.g. on a 32-bit system only about 4GB of memory can be addressed. But even if addressing is not a problem, if the amount of allocated memory exceeds the computer's RAM then swapping occurs which degrades performance and effectively grinds computation to a halt. In practice, this means that no more than a few million points can be tessellated using such implementations.

\begin{figure}[floatfix]
\begin{center}
 \resizebox{8cm}{!} { \includegraphics{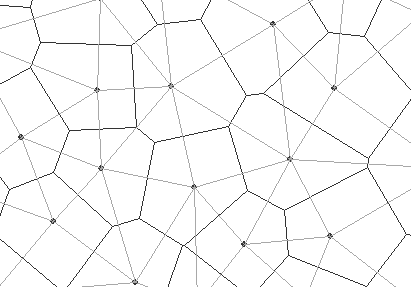} }
\caption{ Sample 2D Delaunay tessellation (gray triangles) and Voronoi partition (black cells) over sites (black dots). }
\end{center}
\end{figure}

One solution is to use shared-nothing distributed computing \cite{Stonebraker}, where each node performs part of the tessellation and the results are joined. In this paper we do not consider distributed computing.

On a single machine either the application performs its own swapping in a way that is optimal or the tessellation process is serialized reducing memory requirements to avoid swapping altogether. Our approach described in this article is of the second kind. Given the current trend in computing to pack many cores into a processor we will comment on the effect of memory-serialization on processor parallelism in the context of Delaunay tessellation.

Section 2 describes existing Delaunay algorithms with some comments on whether memory-serialization is feasible; in Section 3 we present the plane-sweep incremental algorithm. Section 4 gives some implementation details and we briefly describe how we use Delaunay tessellations and Voronoi partitions with astrophysical datasets. Section 5 concludes the paper. 

\section{ Existing Delaunay Algorithms }

We describe algorithms in 2D and cite results whether and how they perform in higher dimensions. For a more complete overview, see \cite{Fortune}, \cite{Cignoni} and \cite{Barber}.

\subsection{ Flipping }
Flipping is bootstrapped by creating some triangulation of $S$ that does not have to conform to the Delaunay condition. The basic idea is that if four sites $a, b, c$, and $d$ are such that the triangles $abc$ and $abd$ are parts of the current triangulation but they are not locally Delaunay, i.e. the circumcircle of $abc$ contains $d$ or vice versa, than the diagonal edge $ab$ is dropped and the new diagonal $cd$ is created; this is called flipping. The new triangles $acd$ and $cbd$ are now locally Delaunay. The algorithm is of $O(n^2)$ time complexity \cite{Fortune}. In higher dimensions, naive flipping does not work \cite{Joe}, however, a derivative, incremental flipping similar to the incremental algorithm described below does \cite{Rajan}. It is interesting to note that flipping is easy to parallelize in 2D, since a simple per triangle locking mechanism can ensure that each triangle is used by one thread only; one thread only works with two triangles at a time (that are not locally Delaunay), and the effects of flipping are local.

\subsection{ Incremental (Bowyer-Watson) }

The incremental algorithm is a direct application of the Delaunay condition. Suppose that a subset of $S$ is already tessellated conforming to the Delaunay condition, and we want to add a new site $s$ to the tessellation. The procedure is to find all triangles in the current tessellation which break the Delaunay condition w.r.t. the new site, i.e. all triangles whose circumcircle contains $s$. Then, remove these triangles from the tessellation producing a convex hole. Finally, connect the edges forming the boundary of the hole to the new site $s$ to form the new triangles which fill the hole. The algorithm trivially generalizes to higher dimensions. Fortune \cite{Fortune} proves that the incremental algorithm is of $O(n^{\ceil{(d+1)/2}})$ time and $O(n^{\ceil{d/2}})$ space complexity. Verma \cite{Verma} shows that the incremental algorithm can be parallelized effectively. To do this, the data structures must be designed to support transactions (create transaction, perform modifications local to the transaction, then commit or rollback). The idea is to perform several insertions simultaneously, then check whether the changes conflict: if they do then roll back some of the transactions, else commit all. Our hybrid approach is based on the incremental algorithm, since the Delaunay condition offers a natural way to serialize it.

\subsection{ Plane-Sweep (Fortune) }
Plane-sweep methods are widely used in computational geometry. Fortune \cite{Fortune} was the first to give a plane-sweep algorithm to construct Voronoi partitions in 2D. The key idea is to move a sweep line (plane in higher dimensions) across the space and maintain a wave front; for sites that have been swept, the set of points equidistant from the line and the site is a parabola. The wave front is the parts of parabolas closest to the sweep line, breaks in the wave front are the intersections of parabolas. The breaks of the wave front trace the edges of the Voronoi cells. The time complexity is $O(n log(n))$ in 2D. The problem with Fortune's approach in higher dimensions is that the geometry of d-dimensional paraboloid intersections gets very complicated, in fact, there is no $d > 3$ implementation. However, plane-sweeping is the obvious approach for constructing a memory serialized algorithm. Our hybrid approach was inspired by Fortune's algorithm, and it also uses a sweep plane.

\subsection{ Divide \& Conquer (DeWall) }
The DeWall algorithm is a divide \& conquer (DC) algorithm which solves the Delaunay tessellation problem in $E^d$ for any $d$. DeWall,separates the sites into two smaller subsets using a cut plane, computes the Delaunay tessellation along the plane (called the wall) and then computes the two disjoint tessellations on the two sides of the wall; no merging is involved. The DeWall algorithm is of $O(n^{\ceil{d/2}+1})$ time and $O(n^{\ceil{d/2}})$ space complexity \cite{Cignoni}. The DeWall algorithm is easy to parallelize: once the wall is constructed the two subsets can be tessellated independently, in parallel.

\subsection{ Convex Hull (QuickHull) }
It is a fundamental result of computational geometry that a convex hull algorithm in $E^{d+1}$ can be used to compute a Delaunay tessellation in $E^d$: elevate the sites in $E^d$ onto a parabola in $E^{d+1}$, compute the convex hull of the elevated points and the projection of the lower envelope of the hull back onto $E^d$ is the Delaunay tessellation. Thus, all convex hull algorithms are candidates for computing Delaunay tessellations. Since these two problems are closely related, many algorithms have analogues in the other domain. Here we will consider the algorithm of the most widely used convex hull implementation, QuickHull. The idea in QuickHull is to start out with any simplex, this is the initial hull. For each face $f$, let $Front(f)$ be the set of sites on the normal side of $f$; for each site $s$, record the $Front$ memberships in $Visible(s)$ ($s$ in is $Front(f)$ iff $f$ is in $Visible(s)$). Given an existing hull, convex hull algorithms select a face $f$ of the hull at random, then select a site $s$ that is in $Front(f)$. Then the horizon ridges of $Visible(s)$ are identified and these are connected to $s$ forming new faces, thus extending the hull. During the process the $Front$ and $Visible$ sets are maintained. The faces inside the old horizon are thrown away. Note that if for a site $s$ the set $Visible(s)$ becomes empty, that means that $s$ is inside the current hull, thus it can be discarded, it will not affect the outcome of the convex hull algorithm. Previous algorithms selected a random site from $Front(f)$ to extend the hull; the idea of QuickHull is to use the site $s$ in $Front(f)$ that is farthest away from $f$. The advantage is obvious: after extending the hull to $s$, hopefully many sites lying closer to the face will be inside the extended hull \cite{Barber}. In practice, the QuickHull algorithm takes much less steps than the randomized version, but in the worst case scenario all strategies are equivalent. In the $d = 2$ case the complexity of QuickHull is $O(n^2)$ if all sites end up lying on the hull, the case if using it to perform Delaunay tessellations. For $d > 2$ the complexity is an open question. Cignoni et. al claim that DeWall is faster than QuickHull \cite{Cignoni}. QuickHull offers good processor parallelization: the hull can be extended at several regions as long as the faces inside the old horizons are disjoint. The standard QuickHull implementation, QHull, performs convex hull and Delaunay tessellations, but it does not support parallel computation and is memory limited.

\section{ Plane-Sweep Incremental Algorithm }
The Plane-Sweep Incremental Algorithm is a simple modification of the incremental algorithm: insert the points in a way that emulates the plane-sweep approach, i.e. insert the points sorted along an axis. The key idea is that once the sweep-line has passed a simplex' circumsphere no new points added later can be contained in the sphere. Thus, this part of the tessellation cannot change and it can be written from memory to disk. We call this offlining.

\subsection{ Improved Offlining Condition }
Let $B$ be the bounding box of all sites. An even better offlining condition for the simplex $T$ with circumsphere $C$ is to compute the maximal $x$ coordinate $x(B, C)$ of the intersection of $B$ and $C$ (assuming the sweep plane moves in the positive $x$ direction). Once the sweep-line has passed $x(B, C)$ no new points added later can be contained in the sphere. This improved offlining condition helps a great deal in reducing memory requirements. 

\subsection{ Search Structure }
The iterative algorithm depends on a search structure to quickly find the simplices whose circumspheres break the Delaunay condition w.r.t. the new site. The idea is to somehow find the seed simplex containing the new point (its circumsphere trivially contains it), then grow the region by adding neighboring simplices whose spheres also contain the point. The search structure is only used to find the seed simplex. In the plane-sweep incremental approach, only simplices near the $d-1$ dimensional sweep plane are online, thus instead of indexing in the original $d$-dimensional space, we index the $d - 1$ dimensional plane. In our implementation, we use a simple high-resolution static grid and found that this is not a bottleneck in the procedure. However, as the plane moves, it sweeps different regions of the point set whose $d-1$ dimensional distribution may be radically different. Thus, a general purpose implementation should use a dynamic kd-tree.

\subsection{ Sweep direction }
The memory requirements of the plane-sweep incremental algorithm are determined by the "width" or "thickness" of the online region following the sweep plane. Thus, "the best sweep direction is parallel to the point set" or in other words "the best sweep plane is perpendicular to the point set". More precisely, the sweep direction should be the first principal axis of the point set. For example if the point set is homogenous in a bounding box with extents $(\Delta x=10, \Delta y=1, \Delta z=1)$, then sweep direction should be $x$.

\subsection{ Memory Usage }
As with the incremental algorithm, the plane-sweep incremental approach is of $O(n^{\ceil{d/2}})$ space complexity in the worst case, meaning that it is possible to construct point sets where little offlining is possible, for example a point set where the points lie on the surface of a sphere. In practice, for real point distributions, we found that the offlining approach works well to limit memory requirements. To illustrate, consider  a homogenous point set with density $\rho$ in the bounding box with extents $(\Delta x > 1, \Delta y=1, \Delta z=1)$. Based on our previous discussion, we choose $x$ as the sweep direction. The nice property of our approach is that the memory usage is independent of $\Delta x$: if the point set consists of $N$ points, then $N \sim \rho \Delta x$ holds, but the number of points in the neighborhood of any $x=const$ section (swept by the plane) relevant to the memory usage is proportional with $(N/\Delta x)^{2/3} \sim \rho^{2/3}$ only. Thus $\Delta x$ affects runtime only.

\subsection{ Runtime Speed }
Due to the set insertion order of points in the plane-sweep incremental approach, for small point distributions (where other algorithms do not run out of memory and fail) other algorithm perform better. In fact, Fortune \cite{Fortune} proves that the plane-sweep approach produces the worst case time complexity of $O(n^{\ceil{(d+1)/2}})$.

\subsection{ Parallelization }

\subsubsection{Parallelization of Voronoi partitioning.}
Computation of Voronoi cell's volumes can be inserted after the simplices are offlined (but before they are written to disk). Since at this point the simplices are already removed from the tessellation we have data independence, thus this task can easily be performed in a parallel thread of execution. The parallelizability here depends on how fast the Delaunay module sweeps, i.e. how fast the points become available for Voronoi computation.

\subsubsection{Parallelization of Delaunay tessellation.}
Verma \cite{Verma} describes a parallel transaction-based Delaunay kernel. The idea is to insert several points at once (parallel operation). Since this operation affects the tessellation only locally, if we're lucky, the affected regions are disjoint, and we can commit all changes (serial operation). If there are conflict regions, we must resolve them (serial operation). Verma reports speedups of 3x. We have not experimented with a parallel Delaunay kernel, so we offer theoretical considerations only. The applicability and effectiveness of this approach depends on the point distribution and insertion order: it will perform best if the points inserted in parallel are far away from each other so that there is no conflict. In practice, this is achieved by inserting points in random order. This is orthogonal to our requirement that points be inserted in a set order, thus, in our case, the effectiveness of this approach depends  only on the point distribution being tessellated. 

\subsection{ Generalization Possibilities }
It is not necessary to emulate a plane, we could emulate other $d-1$ -dimensional submanifolds, like a sphere. When using the plane-sweep incremental algorithm with a new submanifold, two questions have to be answered: (i) what search structure to use (e.g. project onto the submanifold and then use a parameterization of the manifold to create a tree or grid), and (ii) give the algorithm or formula for determining when a simplex can be offlined.

\section{ Application }

\subsection{ Implementation }
We implemented the algorithm in the $d = 3$ special case in C/C++ using Microsoft's Visual C++ compiler. We use some Standard Template Library (STL) data structures, such as std::vector, std::list and std::multimap. Since we are tessellating very large point sets which are stored in a Microsoft SQL Server 2005 database, we directly read/write to the database using bulk copy operations over ODBC. Our implementation contains many optimizations, such as its own memory manager over malloc(). We did not implement Verma's parallel kernel, but we do parallelize the calculation of Voronoi volumes.

\subsection{ Tessellating SDSS Data }
The Sloan Digital Sky Survery (SDSS) is an astrophysical project whose goal is to create a highly detailed map of our universe. Every night, a 2.5 meter telescope photographs the night sky, then sends its data through a processing pipeline. The result is public archive, a database of 287 million celestial objects \cite{SDSS}. The telescope has 5 x 6 CCD arrays, 6 arrays for 5 filters each; the filters are called u, g, r, i, z referring to the part of the electromagnetic spectrum they center on (e.g. r for red). Each of the 287M objects has these 5 magnitudes stored in the database. Our goal was to compute the Delaunay tessellation of this magnitude space. The applications of such a graph are manyfold: in astrophysics it can be used for clustering (identifying different classes of celestial objects) and density estimation of the data distribution (the volume of the Voronoi cell is inversely proportional to the local density). Additionally, a Delaunay-graph explicitly solves the nearest neighbour query problem, a common task in astrophysics.

Our first step was to use principal component analysis (PCA) to transform the data to 3D, which results in little data loss. The resulting distribution is shown in Figure 2. 

\begin{figure}[floatfix]
\begin{center}
 \resizebox{8cm}{!} { \includegraphics{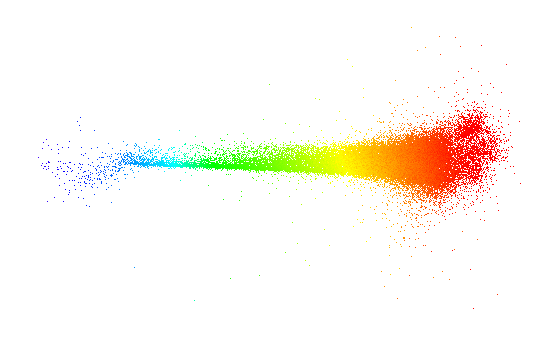} }
\caption{ The SDSS magnitude space; each point corresponds to a celestial object identified by the telescope pipeline. The points are colored according to their first PC coordinate. }
\end{center}
\end{figure}

We chose the first princpial axis as the sweep direction. Additionally, to make the global bounding box narrower, and thus increase offlining, we cut off 4\% of the data at the sides of the distribution. The algorithm's input data is one table, where each row corresponds to a site in Figure 2., and for each site, an id (4 byte integer) and the three coordinates (three 8 byte doubles) are stored. Thus, the table is 8.4GB in size. The program's output is two tables: (i) the table containing the Delaunay graph, where each row corresponds to an edge, and for each edge, two ids (4 byte integer) are stored and (ii) the table containing the Voronoi volumes, where each row corresponds to a site, and for each site, we store the Voronoi cell's volume (8 byte double). The tessellation resulted in 2.1 billion edges, thus the edge table is 17.4GB in size, the Voronoi table is 3.6GB is size. The total running time of the algorithm was 105 hours, while the peak memory usage was only 1.5GB. Figure 3. demonstrates how offlining limited the number of terahedrons that had to be kept in memory.

\begin{figure}[floatfix]
\begin{center}
 \resizebox{8cm}{!} { \includegraphics{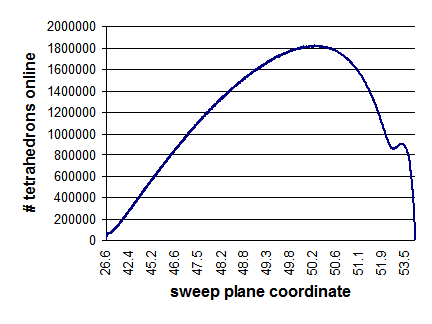} }
\caption{ As the plane sweeps the point cloud, tetrahedrons left behind are offlined, limiting the geometry that has to be kept in memory.  Comparing with Figure 2., the number of online tetrahedrons roughly follows the density of the point distribution. Peak memory usage was 1.5GB. }
\end{center}
\end{figure}

\section{ Conclusion }
In the future, we plan to rewrite our implementation to support any dimension and experiment with a parallel Delaunay kernel. We will investigate the possibility of using a Delaunay graph as a general-purpose database index ideal for nearest neighbour queries, which involves supporting dynamic add() and remove() operations, a topic not considered in this paper.


\begin{thebibliography}{9}

\bibitem{Fortune}
S. Fortune, Voronoi Diagrams and Delaunay Triangulations, Computing in Euclidean Geometry, World Scientific Publishing, p.193-223 (1992).

\bibitem{Stonebraker}
M. Stonebraker, The Case for Shared Nothing Architecture, Database Engineering, Volume 9, Number 1 (1986).

\bibitem{Cignoni}
P. Cignoni, C. Montani, R. Scopigno, DeWall: A Fast Divide \& Conquer Delaunay Triangulation Algorithm in $E^d$, Computer-Aided Design, Volume 30, Number 5, p.333-341 (1998).

\bibitem{Barber}
 C. B. Barber, D. P. Dobkin, H. Huhdanpaa, The Quickhull Algorithm for Convex Hulls, ACM Trans. on Math. Software, Volume 22, Number 4, p.469-483 (1996).

\bibitem{Joe}
B. Joe, Three dimensional triangulations from local transformations, SIAM J. Sci. Stat. Computing, 10:718-741 (1989).

\bibitem{Rajan}
V. T. Rajan, Optimality of the Delaunay triangulation in $R^d$, Proc. Of the Seventh Annual Symp. On Comp. Geom., 357-363 (1991).

\bibitem{Verma}
C. S. Verma, Multithreaded Delaunay Triangulation, Technical Report, Dept. of Computer Science, The College of William and Mary (2004).

\bibitem{SDSS}

J. K. Adelman-McCarthy, et al., The Sixth Data Release of the Sloan Digital Sky Survey, Astrophys. J. Suppl. 175:297-313 (2008).

\end{thebibliography}
\end{document}